\documentclass[aps,pre,floatfix,nofootinbib]{revtex4}
\usepackage{subfigure}
\usepackage{epic}\usepackage{eepic}
\usepackage{bm}
\usepackage[dvips]{epsfig}
\usepackage{graphicx} \usepackage{amsmath} \usepackage{amssymb}
\newcommand{\comment}[1]{}
\renewcommand{\u}{\underline}

\newcommand{\x}{{\bf x}}
\newcommand{\X}{{\bf X}}
\newcommand{\Y}{{\bf Y}}
\newcommand{\y}{{\bf y}}
\newcommand{\n}{{\bf n}}
\newcommand{\xxx}{\boldsymbol{\xi}}
\newcommand{\eee}{\boldsymbol{\eta}}
\newcommand{\eps}{\boldsymbol{\epsilon}}
\newcommand{\et}{\eta}

\renewcommand{\O}{{\rm O}}
\renewcommand{\L}{{\rm L}}
\newcommand{\BEA}{\begin{eqnarray}}
\newcommand{\EEA}{\end{eqnarray}}

\begin{document}

\title{Active image restoration}

\author{Rongrong Xie$^{1,\ast}$, Shengfeng
Deng$^{1,\ast}$\footnotetext[1]{These authors contributed equally to this work.}, Weibing Deng$^{1,\dagger}$ and
Armen E.~Allahverdyan$^{2,\dagger}$\footnotetext[2]{Authors to whom any 
correspondence should be addressed: wdeng@mail.ccnu.edu.cn and armen.allahverdyan@gmail.com}}

\address{$^1$Key Laboratory of Quark and Lepton Physics (MOE) and 
Institute of Particle Physics,  Central China Normal University, 
Wuhan 430079, China \\ $^2$Yerevan Physics Institute, Alikhanian Brothers Street 2, 
Yerevan 375036, Armenia}

\begin{abstract} We study active restoration of noise-corrupted images
generated via the Gibbs probability of an Ising ferromagnet in external
magnetic field. Ferromagnetism accounts for the prior expectation of
data smoothness, i.e. a positive correlation between neighbouring pixels
(Ising spins), while the magnetic field refers to the bias. The
restoration is actively supervised by requesting the true values of
certain pixels after a noisy observation. This additional information
improves restoration of other pixels. The optimal strategy of active
inference is not known for realistic (two-dimensional) images. We
determine this strategy for the mean-field version of the model and show
that it amounts to supervising the values of spins (pixels) that do not
agree with the sign of the average magnetization. The strategy leads to
a transparent analytical expression for the minimal Bayesian risk, and
shows that there is a maximal number of pixels beyond of which the
supervision is useless.  We show numerically that this strategy applies
for two-dimensional images away from the critical regime. Within this
regime the strategy is outperformed by its local (adaptive) version,
which supervises pixels that do not agree with their Bayesian estimate.
We show on transparent examples how active supervising can be essential
in recovering noise-corrupted images and advocate for a wider usage of
active methods in image restoration. 

\end{abstract}

\maketitle 

\section{Introduction}

Many inference problems in machine learning amount to restoration of a
hidden structure based on noisy observations. They are similar
(sometimes isomorphic) to models of equilibrium statistical physics,
where the role of noise is played by frozen disorder. In particular, the
problem of noise-corrupted image restoration can be mapped to the
two-dimensional Ising ferromagnet
\cite{geman,besag,scot,tanaka,cohen,nishimori,nishi}. The Gibbs probability of
this model serves as a prior probability for images.  Spins (two-value
variables equal to $\pm 1$) of the Ising model refer to the black-white
pixels of a digital image. The ferromagnetic feature of the model means
that neighbouring pixels are correlated, a legitimate minimal assumption
on the prior probability. 

Methods of solving inference problems cannot perform without prior
information. Since its amount is limited, it is natural to study
informed strategies for requesting prior information. In particular,
prior information can be requested on the ground of the previous
functioning of the inference method. This notion of active
inference~\cite{review_active} emerged in statistics \cite{wald} and has
been applied to various inference problems including Hidden Markov
Models~\cite{Anderson2005,Hoffman2001,ag_active}, network
inference~\cite{Zhu2003,Bilgic2009,Bilgic2010,Moore2011}, economics,
experiment optimization {\it etc} \cite{review_active}.  In the context
of noisy image restoration, the concept of active inference assumes that
there is a costly noiseless channel by which the correct values for some
of pixels can be communicated directly to the image restorator. In
practice, this noiseless channel may mean an additional precise (but
costly) measurement of pixels. Since the channels is costly, only a
small fraction (say $5\%$) of original pixels is communicated with a
hope that|if employed properly|they can lead to a sizable improvement in
restoration of other (noisy) pixels.  The subject of active inference
has two related issues: which prior information is to be requested and
to what extent this information improves the method performance
\cite{review_active}. The first question refers to the meaning of (prior)
information, the second one to its value. 

Most existing theoretical results on active inference in graphical
models are concerned with sub-modular cost functions that allow to
establish certain optimality
guarantees~\cite{Krause2005_UAI,Krause2009_JAIR}. The sub-modularity
assumption does not hold in several interesting situations, e.g.  for
systems with long-range correlations and scale-invariance, a range of
phenomena that is collectively designated as critical behavior. The
clearest example of such a behavior is the second-order phase
transitions in statistical mechanics \cite{nishi}. It was found that
natural images (e.g. nature scenes) are organized in a scale-invariant
way \cite{ruderman} and do have long-range correlations \cite{bialek}.
Statistical mechanics (e.g.  the Ising model) is the most appropriate
way of describing such images \cite{bialek}. Generally, the optimal
strategy of active inference|i.e.  which variables are to be supervised
given a noisy observation|is not known beyond a direct enumeration
\cite{Krause2009_JAIR}. 

Here we shall study an active recognition (estimation or filtering) of a
noise-corrupt image, which was generated via the two-dimensional Ising
ferromagnet. The model does allow for a critical regime depending on the
value of the ferromagnetic coupling constant, which is taken as the
known hyperparameter of the model. For a large number of pixels, this
critical regime is realized via a second-order phase-transition, with
magnetization being an order parameter. 

We analytically determined the optimal active supervising strategy of
the Ising model in the mean-field limit. This mean-field limit of the
Ising model was already studied for a Gaussian noise image-restoration
problem \cite{nishi,nishimori}.  More generally, mean-field methods are
widely applied in statistical inference problems
\cite{jung,forbes,inoue,geiger,mitter}; see \cite{opper} for a book
presentation.  The inferred image is determined by minimizing the
Bayesian risk \cite{robert}. The optimal mean-field strategy amounts to
supervising only those pixels whose observed (noisy) value does not
agree with the sign of the magnetization.  It also shows that there is a
maximal number of supervised variables, beyond of which the supervising
is not meaningful.  Without active supervising, the optimal inference of
the mean-field model shows only two (in a sense trivial) regimes, {\it
viz.} observation-dominated (where the prior information is not needed)
and prior-dominated (where observations are redundant).  With active
supervising, there is a non-trivial coupling between observations and
the prior knowledge, i.e.  no separation into two different regimes is
possible. 

The mean-field solution of the actively supervised model is transparent
and guides the understanding of a more complex, two-dimensional
situation which is studied numerically. We saw that the mean-field
optimal active supervising strategy performs well in the two-dimensional
case, if the model is away from the critical regime. Otherwise, we found
a strategy that is superior to the mean-field optimal one. This strategy
amounts to comparing the observed (noisy) value of pixel with its
Bayesian estimate: only those pixels are supervised for which these
values disagree. We show on concrete examples how usual (not active)
methods fail to restore noise-corrupted images, while the active
supervising does lead to a satisfactory restoration.

\comment{For realistic massive models (a large number of estimated
variables) recognition cannot be carried out in a closed form. Hence one
has to rely on various approximation and/or surrogate methods, e.g. the
EM (expectation-maximization) method for hyperparameter estimation.
Since we want to gain a conceptual undersanding of active learning, we
work with a model, where there is a way to determine the optimal active
recognition method analytically. 

The second objective {\it (ii)} is there as well, since the model
allows for the second-order phase transition critical regime with
respect to the (prior) magnitude of the inter-pixel interaction. As
for {\it (iii)}, we noted that without active methods the optimal
filtering of the model shows only two (in a sense trivial) regimes,
{\it viz.}  observation-dominated (where the prior information is not
needed) and prior-dominated (where observations are redundant); see
section \ref{ising}. In addition, no reliable estimation of
hyperparameters is possible, i.e. we have (without active methods) a
relatively strong example of the non-identifiability problem with
respect to hyperparameters \cite{1957,ito}. We show that in this
situation, the active methods do indeed perform well: they improve the
filtering quality via an interplay betweem the noise and prior
information, and (simultaneously) they do allow a complete recovery of
the hyperparameters. Moreover, the active filtering regime allows for
an interpretation in terms of verification and falsifiability. These features of
supervising can hold for more general models.}

This text is organized as follows. Section \ref{gen} recalls the general
theory of optimal Bayesian inference with and without supervising. Here
we also recall how this theory can be represented via Ising models.
Next section finds out the optimal active supervising strategy in the
mean-field version of the Ising model. Section \ref{numo} studies
numerically the two-dimensional Ising model. It shows that several
results deduced in the mean-field situation also apply in this more
realistic case. Section \ref{numo} also works out a local version of
that optimal supervising strategy and shows that it yields better
results in the critical regime of the model, where the mean-field method
does not apply. Section \ref{magno} studies the Ising model, where the
prior probability has a small bias (in the value of spins) given by a
weak external magnetic field.  Note that sections \ref{numo} and
\ref{magno} (numerical results on a square Ising lattice) can be read
independently from \ref{grz} (analytic results in the mean-field limit). 

We summarize in the last section. In
preparation for future research, Appendix studies hyperparameter
estimation of the mean-field Ising model. The analysis here is deeper
than normally done in literature, because we study the influence of
non-identifiability on the optimal restoration. 

\section{General formulation of image restoration and
representation via Ising models}

\label{gen}

\subsection{Without supervising}

Here we recall general ideas of the Bayesian inference for two-valued
random variables with and without supervising. Next sections will
specify them for the image restoration problem. 

Let there are two dependent vector random variables $\X=(X_1,...,X_N)$
and $\Y=(Y_1,...,Y_N)$. Now $\X$ is called the hidden variable, since it
cannot be observed directly. Instead, we assume that a concrete value
$\y=(y_1,...,y_N)$ of $\Y$ is observed. On the ground of $\y$ we want to
find a representative value (an estimate) $\xxx(\y)$ of
$\x=(x_1,...,x_N)$ that is likely to be the source of $\y$. We shall
assume that $y_k=\pm 1$, $x_k=\pm 1$ and call them spins (pixels).  The
joint probability is $P(\X,\Y)$; e.g. the conditional probability
$P(\Y|\X)$ describes a noisy channel, where $\Y$ is the noise-corrupted
observation of $\X$. 

The quality of the estimate $\xxx(\y)$ is given by the average Hamming
distance (or overlap):
\begin{eqnarray}
  \label{eq:14}
  \O (\y;\xxx )
=\frac{1}{N}\sum_{i=1}^N \sum_\x \xi_i(\y)x_i P(\x|\y), 
\end{eqnarray}
so that a larger $\O(\y;\xxx)$ means a better restoration. We stress that
the averaging over $P(\x|\y)$ in (\ref{eq:14}) is done ``by hands'': since
the hidden variables are not known, one generates them via $P(\x|\y)$
(for fixed observations) and then takes the average. Eq.~(\ref{eq:14})
coincides with the inverse Bayesian risk \cite{robert}. 
The latter is minimized, while
(\ref{eq:14}) is maximized over $\xi_i(\y)$.

In applications one also frequently studies the average of $\O(\y)$ over
$P(\y)$. This averaging can also be done by hands, i.e. over
sufficiently many applications, but it is important to stress that if
$N$ is sufficiently large (and $\Y$ is an ergodic process), $\O
(\y;\xxx)$ does self-average, i.e. according to the law of large numbers
the sum (\ref{eq:14}) over a large number $N\gg 1$ is replaced by its
average $\O (\y;\xxx) \simeq \sum_{\y}P(\y)\O (\y;\xxx)$. 

Maximizing (\ref{eq:14}) over $\xxx(\y)$ leads to the optimal method
($i=1,...,N$) with the largest overlap $\hat \O(\y)$ for given
observations $\y$:
\begin{eqnarray}
  \label{eq:5}
&& \hat \xi_i(\y)={\rm sign}\left[\sum_\x x_i P(\x|\y)  \right]
={\rm sign}\left[\,\sum_{x_i=\pm 1} x_i P(x_i|\y)  \right], \\
&& \hat  \O (\y)=\frac{1}{N}\sum_{i=1}^N \left|\sum_\x x_i P(\x|\y)\right|.
  \label{eq:15}
\end{eqnarray}
Eqs.~(\ref{eq:15}) can serve, e.g. for evaluating the
maximum-likelihood method (ML). Recall that the ML does not employ the
prior probability $P(\x)$, and is based on maximizing $P(\y|\x)$ for
given observations $\y$:
\begin{eqnarray}
  \label{eq:16}
  \xxx^{\rm ML } (\y)={\rm arg\,\, max}_{\x}[\,P(\y|\x)\,]. 
\end{eqnarray}
If the prior probability $P(\x)$ is available, $\xxx^{\rm ML } (\y)$ is
suboptimal from the viewpoint of (\ref{eq:5}), and one can determine
how far its prediction for the overlap is from the optimal value $\O
(\y)$ given by (\ref{eq:15}).

For completeness we mention an equivalent approach to introducing the
overlap (\ref{eq:14}). Let us temporarily assume that we know the
original realization $\x$ of the random variable ${\bf X}$.  Once it
passes through the noisy channel, we obtain $\y=\eps\,\x$, where $\eps$
is the noise variable and $\eps\,\x$ means elemetwise multiplication of
two vectors (Hadamard product).  Now the overlap can be defined as
\begin{eqnarray}
\label{pseudo}
Q=\frac{1}{N}\sum_{i=1}^N x_i\hat{\xi}_i(\eps\,\x). 
\end{eqnarray}
The joint distribution
of $x$ and $\eps$ is $P_{{\bf X}{\bf Y}}(\x,\eps \x)=P(\x,\eps \x)$.  

To make $Q$ independent from the concrete choice of $\x$ and $\eps$
we look at the average
\begin{eqnarray}
\overline{Q}=\frac{1}{N}\sum_{i=1}^N\sum_{\x,\, \eps}P(\x,\eps \x) x_i\hat{\xi}_i(\eps\,\x)
=\sum_{\y}P(\y) \hat\O (\y), 
\label{saud}
\end{eqnarray}
which brings us back to the averaged (\ref{eq:15}). However, there is no direct relation
between $\O (\y)$ and $Q$.

\subsection{With supervising}
\label{supero}

Given observations $\y$ one requests an additional information on
certain $x_i$ (supervising) so as to improve the quality of restoration
for the remaining variables. This is described by the vector
\begin{eqnarray}
  \label{eq:29}
  \n=(n_1,...,n_N), \qquad n_i=0,1,
\end{eqnarray}
where $n_i=1$ ($n_i=0$) means that the true value of the corresponding
$x_i$ is requested (not requested). Naturally, the number of
supervised spins is fixed (i.e. not all spins are supervised)
\begin{eqnarray}
  \label{eq:30}
  \sum_{i=1}^N n_i=\rho N. 
\end{eqnarray}
The supervising strategy is described by conditional probability
$P(\n|\y)$, i.e. the supervising is generally probabilistic. Let us
divide $\x$ into supervised ($\x''$) and not supervised ($\x'$)
parts. $\x'$ and $\x''$ depend on $\n$, but this dependence is not
indicated explicitly to avoid excessive notations.  E.g. if
$\x=(x_1,x_2,x_3)$ and $\n=(1,0,0)$, then $\x'=(x_2,x_3)$ and
$\x''=(x_1)$. Let $P(\x',\y)$ and $P(\x'',\y)$ be (respectively) the
joint probability of non-supervised and supervised spins and
observations. These probabilities are found from $P(\x,\y)$.

Let the values of $\x''$ were requested and send via a noiseless
channel (we explain below the meaning of this assumption). 
They appeared to be $\x''=\eee$. Together with $\y$ and $\n$,
also $\eee$ is by now given. Hence the distribution $P(\x'|\eee,\y)$
of the non-supervised spins is conditioned both by observations $\y$
and by supervised variables $\eee$.

We estimate non-supervised spins as
$\xxx'(\eee,\y,\n)$.  The corresponding overlap reads:
\begin{eqnarray}
  \label{eq:a14}
  \O (\eee,\y,\n;\xxx')
=\frac{1}{N(1-\rho)}
\sum_{k=1}^N (1-n_k) \sum_{\x'} \xi'_{k}(\eee,\y,\n)\,x'_k\, 
P(\x'|\eee,\y), 
\end{eqnarray}
where the summation is taken over non-supervised spins only. For a
given $\n$, the optimal overlap is calculated as in (\ref{eq:5},
\ref{eq:15}):
\begin{eqnarray}
  \label{eq:a50}
\hat \xi'_{k}(\eee,\y,\n)&=&{\rm sign}\left[\sum_{\x'} x'_k P(\x'|\eee,\y)
\right]={\rm sign}\left[\sum_{x'_k=\pm 1} x'_k P(x'_k|\eee,\y)
\right],  \\
  \label{eq:a150}
  \hat  \O (\eee,\y,\n)&=&
\frac{1}{N(1-\rho)} \sum_{k=1}^N (1-n_k)\left| \sum_{x'_k=\pm
  1}x'_kP(x'_k|\eee,\y)\right|. 
\end{eqnarray}
Now if $\y$, $\n$ and $\eee$ are sufficiently long, $\hat \O
(\eee,\y,\n)$ will self-average over
$P(\eee,\y,\n)=P(\y)P(\eee|\y)P(\n|\y)$.

The major problem of supervising is to find the best $P(\n|\y)$, which
holds (\ref{eq:30}) and provides the largest average overlap
(\ref{eq:a150}). For understanding this problem one should rely on
models, because there is no general solution for it under $N\gg 1$ (for
a small value of $N$ one can proceed with straightforward calculations). 

Let us now explain the practical meaning of the above assumption on the
existence of a noiseless channel. We note that frequently such channels
do exist (e.g. because they relate to a sufficiently precise equipment), 
but they are costly, i.e. the cost per pixel (or spin) for
using such a channel is high. In that case it is useful to employ the
noisy channel for $N$ pixels, but still to use the noiseless channel for
$\rho N$ actively supervised pixels, where $\rho\ll 1$. 

\comment{
One particular case of the above contruction is worth mentioned: if
$\xi'_{k}(\y,\n)$ in (\ref{eq:a14}) does not depend on $\eee$, and $\O
(\eee,\y,\n;\xxx')$ self-averages over $P(\eee|\y)$, we get
\begin{eqnarray}
  \label{eq:14}
  \O (\y,\n,\xxx')
=\frac{1}{N(1-\rho)}
{\sum_{k}}' \sum_{\x'} \xi'_{k}(\y,n)\,x'_k\, 
P(\x'|\y).
\end{eqnarray}
Maximizing over $\xi'_{k}(\y,n)$ we get 
as in (\ref{eq:a50}, \ref{eq:a150}):
\begin{eqnarray}
  \label{eq:50}
\hat x'_{k}(\y,\n)&=&{\rm sign}\left[\sum_{\x'} x'_k P(\x'|\y)
\right],  \\
  \label{eq:150}
\hat  \O (\y,\n)&=&\frac{1}{N(1-\rho)}
{\sum_{k}}' \left|\sum_{\x'} x'_i P(\x'|\y)\right|\\
  \label{or}
&=&\frac{1}{N(1-\rho)}
\sum_{i=1}^N \left|\sum_{\x} (1-n_i)x_i P(\x|\y)\right|.
\end{eqnarray}
Now $\hat  \O (\y,\n)$ can also be averaged over $P(\y)P(\n|\y)$. This
can also happen via the self-averaging scenario, as we metioned
above. 
}

\subsection{Ising model}
\label{ising}

Let us now assume that the prior probability of hidden variables
$\x=(x_1,...,x_N)$ is given as
\cite{geman,besag,scot,tanaka,cohen,nishi}
\begin{eqnarray}
  \label{eq:1}
  P(\x)\propto e^{{J'}\sum_{\{i,k\}}x_ix_k}, \qquad x_i=\pm 1,
\end{eqnarray}
where $J'>0$ is a constant hyperparameter (coupling constant), and
${-J'}\sum_{\{i,k\}}x_ix_k$ is the Hamiltonian of the Ising ferromagnet,
where $\{i,k\}$ means summation over neighbours of a lattice. In the
context of the image recognition problem this is a square lattice.  Here
$x_i=\pm 1$ refer to black-white pixels of the original image, and the
ferromagnetic coupling $J'>0$ means that there is a prior information on
positive pixel-pixel correlations.  This is just the standard smoothness
assumption. 

We assume that the observed variables $\y=(y_1,...,y_N)$
(noise-corrupted image) relate to $\x$ via a symmetric and independent
noise
\begin{gather}
  \label{eq:2}
  P(\y|\x)=\prod_{i=1}^N p(y_i|x_i)=(2\cosh[h])^{-N}\prod_{i=1}^N
e^{h\sum_i x_iy_i},\\
 h\equiv\frac{1}{2}\ln\frac{1-\epsilon}{\epsilon}\geq 0,
\label{calvin}
\end{gather}
where $h$ relates to the error probability $0<\epsilon<1/2$, and where we
recall that $y_i=\pm 1$. 

Combining (\ref{eq:1}) and (\ref{eq:2}) we get for the joint
probability $P(\x,\y)$ of hidden variables $\x$ and observations $\y$
\begin{eqnarray}
  \label{eq:3}
  P(\x,\y)\propto e^{-H(\x,\y)}, \qquad 
  H(\x,\y)= - {J'}\sum_{\{i,k\}}x_ix_k -h\sum_i x_iy_i ,
\end{eqnarray}
Now $P(\x,\y)$ refers to the Gibbs distribution of a statistical system
at temperature $T=1$ and with Hamiltonian $H(\x,\y)$. We shall assume
that $J$ and $h$ are known hyperparameters. More generally, they are not
known, but should be inferred from data; see \cite{berg} for a recent 
review. Standard methods for doing that
are discussed in Appendix together with their limitations.

\section{Mean-field analysis}
\label{grz}

\subsection{The fully coupled (mean-field) Ising model}
\label{mf}

To get from (\ref{eq:3}) a solvable model, we assume in (\ref{eq:1}, \ref{eq:3}) 
that all $x_i$ couple with each other:
\BEA
\label{eq:77}
H(\x,\y)= - \frac{J}{2N}\sum_{i\not =k}x_ix_k -h\sum_i x_iy_i ,
\EEA
where we also assumed $J'=J/(2N)$ to show that we have to have
$H(\x,\y)={\cal O}(N)$ for $N\to\infty$. The first sum in (\ref{eq:77})
goes through all $i,j=1,...,N$ under $i\not = j$. Eq.~(\ref{eq:77})
refers to the mean-field-interaction version of the random-field Ising
model. It is known to be solvable \cite{nishi,nishimori}. 

\comment{Hence we apply above general formulas to a concrete model. We choose
to work with (perhaps) the simplest possible model which has $N\gg 1$
interacting variables, and which (simultaneoulsy) does make necessary
to apply supervising, since its non-supervised performance is poor.}

Eqs.~(\ref{eq:3}, \ref{eq:77}) lead to
\begin{eqnarray}
\label{eq:8.0}
P(\x,\y)
\propto
e^{\frac{J}{2N}\sum_{i\not =k}x_ix_k +h\sum_i x_iy_i}
=\frac{N^{3/2}}{\sqrt{2\pi  J}}\int {\rm d} \mu \exp\left[
-\frac{N J \mu^2}{2}+\sum_i x_i(J\mu+hy_i)
\right],\\
\label{eq:8}
P(\y)
\propto
\sum_x e^{\frac{J}{2N}\sum_{i\not =k}x_ix_k +h\sum_i x_iy_i}
=\frac{N^{3/2}}{\sqrt{2\pi  J}}\int {\rm d} \mu \exp\left[
-\frac{N J \mu^2}{2}+\sum_i \ln\cosh(J\mu+hy_i)
\right].
\end{eqnarray}
For $N\gg 1$ the latter integral is taken by the saddle-point
method:
\begin{eqnarray}
  \label{eq:9}
  P(\y)  \simeq \exp\left[
    -\frac{N J m^2}{2}+\sum_i \ln\cosh(Jm+hy_i)
  \right], 
\end{eqnarray}
where the magnetization $m$ is determined from the saddle-point equation as 
\begin{eqnarray}
  m=\frac{1}{N}\sum_i \tanh(Jm+hy_i).
  \label{eq:9.1}
\end{eqnarray}
Now formally $m$ is a function of $\y$, but for this model (and in the
limit $N\gg 1$) it self-averages and becomes (almost) independent from
$\y$. This known fact can be confirmed via calculating correlation
functions between $(x_i,y_i)$ and $(x_j,y_j)$. Thus we return to 
(\ref{eq:8.0}) employ there the saddle-point method, use
(\ref{eq:9.1}) and end up with
\begin{eqnarray}
  \label{rashid}
&&  P(\x,\y)\simeq\prod_i \pi (x_i,y_i), \qquad \pi (x,y)= 
\frac{e^{J m x +h x y}}{2\cosh[Jm+h]+2\cosh[Jm-h]},\\
  \label{eq:10}
&&  P(\y)\simeq\prod_i \pi (y_i), \qquad \pi (y)=\frac{ \cosh (Jm+hy) }
{\cosh[Jm+h]+\cosh[Jm-h]}=\frac{1}{2}[1+y\tanh(Jm)\tanh(h)],
\end{eqnarray}
where $m$ is to be determined from (\ref{eq:9.1}) via self-averaging:
\begin{eqnarray}
  \label{eq:11}
  m=\sum_{y=\pm 1} \pi(y) \tanh(Jm+h y)=\tanh(Jm). 
\end{eqnarray}
One can verify from $P(\x)$ obtained via (\ref{eq:8.0}) that $m$ coincides with 
the average collective spin of the original image: $m=\frac{1}{N}\sum_\x P(\x)\sum_i x_i$.

For $J<1$, (\ref{eq:11}) predicts $m=0$. For $J>1$, (\ref{eq:11})
predicts two solutions with $m>0$ and $m<0$. They refer to two
different ergodic components. Thus we have a second-order phase
transition at $J=1$. It belongs to the mean-field universality class.

\comment{If $\Y$ is an ergodic process, (\ref{eq:15}) self-averages in the
limit $N\gg 1$:
\begin{eqnarray}
  \label{eq:7}
  \O(\y)\simeq \O\equiv \sum_\y \O(\y) P(\y).
\end{eqnarray}
Otherwise, the averaging in (\ref{eq:7}) should be restricted to one
ergodic component.}

Thus (\ref{eq:5}, \ref{rashid}, \ref{eq:10}) imply for the optimal
situation
\begin{eqnarray}
  \label{eq:12}
  \hat \xi_i (y)={\rm sign}[Jm+hy_i],\qquad i=1,...,N,
\end{eqnarray}
and the self-averaged optimal overlap reads from (\ref{eq:15})
\begin{eqnarray}
  \label{eq:13}
\hat  \O=\sum_{y=\pm 1} \pi(y) \tanh(|Jm+hy|)
  ={\rm max}[\,\tanh(J|m|), \tanh(h)\,]. 
\end{eqnarray}
Note that $\tanh(h)$ is the overlap of the maximum-likelihood method;
see (\ref{eq:16}, \ref{eq:2}). This is confirmed by taking $\xi_i^{\rm
  ML} ={\rm sign}[y_i]$ and using it together with (\ref{eq:10}) in 
\begin{eqnarray}
\O^{\rm ML}=\sum_{y=\pm 1} \pi(y){\rm sign}[y] \tanh(Jm+hy)
=\tanh(h).
\label{kio}
\end{eqnarray}

Hence the message of (\ref{eq:12}, \ref{eq:13}) is that there are only
two extreme situations: for $h>J|m|$|which means a weak noise according
to (\ref{calvin})|the prior information is irrelevant, since
observations are reliable. Hence the optimal estimation method coincides
with the maximum-likelihood; see (\ref{eq:16}, \ref{eq:2}). For $h<J|m|$
(strong noise, as seen from (\ref{calvin})) observations are irrelevant,
since the estimate (\ref{eq:12}) depends only on the parameter $J$ of
the prior $P(\x)$. Put differently, either the prior information (given by $J$
in $P(\x)$) is irrelevant, or observations are irrelevant.

\subsection{Supervising}
\label{super}

\subsubsection{The optimal overlap after supervising}

We recall from section \ref{supero} that supervising is described by
the conditional probability $P(\x'|\eee,\y)$ of non-supervised spins
given the values of supervised spins $\eee$ and observations $\y$, as
well as by the conditional probability $P(\n|\y)$ of the coordinates
$\n$ of supervised spins. Now (\ref{rashid}, \ref{eq:10}) show that
for the present model the probabilities $P(\x,\y)$ and $P(\y)$
factorize. This implies that $P(\x'|\eee,\y)$ does not depend on
$\eee$. Using again (\ref{rashid}, \ref{eq:10}) we get from 
(\ref{eq:a150}):
\begin{eqnarray}
  \label{eq:23}
  \hat  \O (\eee,\y,\n)=  
\frac{1}{N(1-\rho)} \sum_{k=1}^N (1-n_k)\tanh[|Jm+hy_k|],
\end{eqnarray}
where $m$ is determined from (\ref{eq:11}). Hence the average of
(\ref{eq:23}) over $P(\n|\y)P(\y)$ reads
\begin{eqnarray}
\label{eq:9.4}
\hat{\O}
=\frac{1}{1-\rho}
\sum_{y=\pm 1}\pi(y)p(0|y)\tanh[|Jm+hy|],
\end{eqnarray}
where $\pi(y)$ is found from (\ref{eq:10}), and where we assumed that
$P(\n|\y)$ (the probability of $\n$, given the observations $\y$) is
symmetric in the sense that all $(y_i,n_i)$ do have the same
marginal-conditional probability: $p(n_i|y_i)$.  The constraint
(\ref{eq:30}) will be implemented in average.  Hence $p(n|y)$ holds
two constraints:
\begin{eqnarray}
\label{eq:9.11}
&& 1=p(1|y)+p(0|y), \\
\label{eq:9.2}
&&1- \rho= \sum_{y=\pm 1}\pi(y)p(0|y).
\end{eqnarray}
The meaning of (\ref{eq:9.4}) is intuitively clear: the optimal
supervising amounts to changing|from $\pi(y)$ to $\pi(y)p(0|y)$|the
(effective) distribution of observations, where $p(0|y)$ is the
probability of not supervising a given spin. The same idea can be
implemented as an anzatz for more general models, but there it does not have to
be optimal.

\subsubsection{Random supervising  provides no advantage}

Let us first consider the random supervising, where $p(n|y)$ does not
depend on $y$. This implies from (\ref{eq:9.2}):
\begin{eqnarray}
  \label{eq:35}
  p(0)=1-\rho, \qquad p(1)=\rho. 
\end{eqnarray}
It should be clear that we get from (\ref{eq:9.4}) the same expression
(\ref{eq:13}) as without any supervising.

\subsubsection{Active supervising}
  
We turn to the active supervising and maximize $\hat{\O}$ given by
(\ref{eq:9.4}) over $p(0|y)$ under constraints (\ref{eq:9.11},
\ref{eq:9.2}). To understand the idea of maximization, assume $m>0$.
Then $|Jm+h|=Jm+h>|Jm-h|$, and (\ref{eq:9.4}) maximizes upon taking
$p(0|1)$ such that $\pi(1)p(0|1)$ is maximally large and compatible with
(\ref{eq:9.11}), and after that taking $p(0|-1)$ as large as constraints
(\ref{eq:9.11}, \ref{eq:9.2}) still allow. The general solution is
written via introducing 
\begin{eqnarray}
  \label{tupo}
\zeta=\frac{1}{2}[1+\tanh(J |m|)\tanh(h)],
\end{eqnarray}
which is equal to $\pi(1)$ if $m>0$. Then the probabilities maximizing 
(\ref{eq:9.4}) read
\begin{eqnarray}
  \label{guppi1}
  \label{yu}
&& {\rm for}\quad m>0 \qquad p(0|1)={\rm min}\left[\frac{1-\rho}{\zeta}, 1\right], \qquad p(0|-1)={\rm max}\left[0,1-\frac{\rho}{1-\zeta}\right], \\ 
&& {\rm for}\quad m<0 \qquad p(0|-1)={\rm min}\left[\frac{1-\rho}{\zeta}, 1\right], \qquad p(0|1)={\rm max}\left[0,1-\frac{\rho}{1-\zeta}\right].
  \label{guppi2}
  \label{la}
\end{eqnarray}
Note that in (\ref{guppi1}) the first (second) argument of min is
selected together with the first (second) argument of max. The same
holds in (\ref{guppi2}). The optimal $\hat{\O}$ reads from 
(\ref{guppi1}, \ref{guppi2}):
\begin{eqnarray}
\label{eq:40}
\hat{\O}={\rm min}\left[\frac{\zeta\tanh[J|m|+h]+(1-\rho-\zeta)
\tanh[\,|\,J|m|-h|\,]}{1-\rho}, ~ \tanh [\,J|m|+h]\, \right].
\end{eqnarray}
When $\rho$ starts to increase from $\rho=0$, (\ref{eq:40}) monotonously
increases from its $\rho=0$ value given by (\ref{eq:13}) till its
maximal value $\tanh [\,J|m|+h]$. According to the non-supervised
maximal overlap (\ref{eq:13}), for $J>1$, but $Jm<h$ (i.e. the noise is
weak, as seen from (\ref{calvin})) the prior is not relevant. This is improved after
supervising, now the prior is always relevant provided that $J>1$. It is
not meaningful to supervise for $1-\rho<\zeta$, since having reached
$\tanh [\,J|m|+h]$, the overlap there does not anymore increase with
increasing $\rho$; see (\ref{eq:40}). Hence the values of 
$p(0|1)=\frac{1-\rho}{\zeta}$ and $p(0|-1)=0$ in (\ref{yu})|as well as
$p(0|-1)=\frac{1-\rho}{\zeta}$ and $p(0|1)=0$ in (\ref{la})|are redundant.
Then (\ref{yu}, \ref{la}) show that observations that agree with the sign of the average 
magnetization $m$ should not be supervised: $p(1|{\rm sign} [m])=0$.

So far we assumed that the hyperparameters $J$ and $h$ in (respectively)
$P(\x)$ and $P(\y|\x)$ are known precisely. Generally, this is not the
case, and hyperparameters themselves are to be found from data; 
see \cite{berg} for a recent review. Appendix
discusses this problem for the considered model.

\section{Numerical results for a square lattice}
\label{numo}

\subsection{Two methods for active supervising}

\begin{figure*}[htbp]
\centering
    \includegraphics[width=0.45\columnwidth]{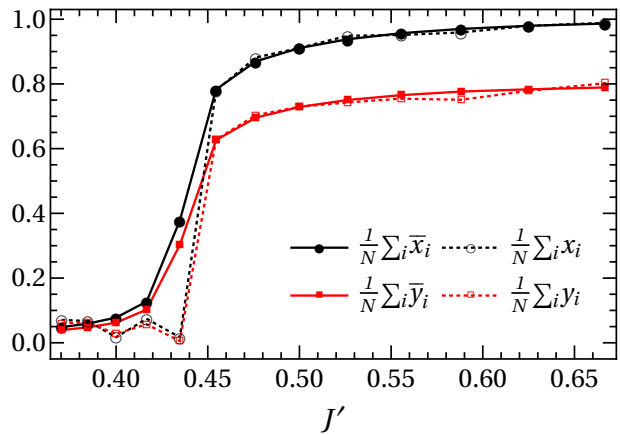}
\caption{ Magnetization $\frac{1}{N}\sum_{i=1}^N{x}_i$ and its
observed value $\frac{1}{N}\sum_{i=1}^N{y}_i$ versus $J'$ under a fixed
$\epsilon=0.1$ in (\ref{eq:2}) (hence $h=1.09861$). For each $J'$ single
realizations of $\x=\{x_i\}_{i=1}^N$ and $\y=\{y_i\}_{i=1}^N$ were
generated according to (\ref{eq:1}, \ref{eq:2}) (i.e. the external magnetic 
field is zero). We also plotted the
averaged values $\frac{1}{N}\sum_{i=1}^N\overline{x}_i$ and
$\frac{1}{N}\sum_{i=1}^N\overline{y}_i$ under the same $\epsilon=0.1$.
For each fixed $J'$ the averaging was taken over 1000 realizations of
random variables $\x$ generated according to (\ref{eq:1}, \ref{eq:2})
and selected such that $\frac{1}{N}\sum_{i=1}^N{x}_i>0$ for all samples
included in the averaging. It is seen that
$\frac{1}{N}\sum_{i=1}^N{x}_i$ and $\frac{1}{N}\sum_{i=1}^N{y}_i$
increase sharply in the vicinity of $J'=0.45$ indicating that in the
limit $N\gg 1$ the system has a second order phase-transition; cf. 
the discussion in the beginning of section \ref{resu}. }
\label{f1}
\end{figure*}

The above results concerned the mean-field model, i.e. an unrealistic
situation if we take into account that real images are two-dimensional.
Hence we turn to studying numerically the Ising model (\ref{eq:3}) on a
square lattice with periodic boundary conditions. 
We start from (\ref{eq:a14}) and (\ref{eq:3}),
and study two different strategies of active supervising. In the first
(global) strategy we supervise $N\rho$ spins selecting them via the
following criterion: given the observation vector $\y$, the $N\rho$
supervised spins $\eta_i$ are chosen randomly among those that hold
\BEA
\label{glo}
y_i\,{\rm sign}\left[\frac{1}{N}\sum_{i=1}^Ny_i\right]=-1,
\EEA
i.e. those spins are supervised which do not agree with the sign of the
collective value $\frac{1}{N}\sum_{i=1}^Ny_i$. Thus variables $\n$ 
in (\ref{eq:29}) are determined via (\ref{glo}). This strategy is clearly
inspired by the above mean-field solution (\ref{guppi1}, \ref{guppi2}),
where $\frac{1}{N}\sum_{i=1}^Ny_i$ is the observed magnetization; see
Fig.~\ref{f1} for the value of $\frac{1}{N}\sum_{i=1}^Ny_i$ averaged
over many samples, as well as a single-sample form of it. For
$\epsilon<1/2$ (which we assume to be the case from (\ref{eq:2})) and a
sufficiently large $N$, we get that $\frac{1}{N}\sum_{i=1}^N x_i$ and
$\frac{1}{N}\sum_{i=1}^Ny_i$ have the same sign.  Note that strategy
(\ref{glo}) is easy to implement. 

Within the second (local) strategy we randomly select $N\rho$ supervised
spins among those that hold
\BEA
\label{lo}
y_i\,\hat{\xi}_i(\y)=-1,
\EEA
i.e. now one first calculates the optimal estimate $\hat{\xi}_i(\y)$
according to (\ref{eq:5}, \ref{eq:3}) and then supervises those spins
that do not agree with observations. This strategy did not show up
within the optimal solution for the mean-field situation.  We choose it
with an expectation that can work, where the mean-field method does not apply,
i.e. fluctuations are essential. 

\comment{
Note that (\ref{lo}) is indeed a local version of (\ref{glo}), as
follows from (\ref{eq:5}) for $\hat{\xi}_i(\y)$ and from noting that for
$N\gg 1$, sufficiently large $J'$ and $\epsilon<1/2$ we have ${\rm
sign}\left[\frac{1}{N}\sum_{i=1}^Ny_i\right]={\rm
sign}\left[\frac{1}{N}\sum_{i=1}^Nx_i\right]$ and
$\frac{1}{N}\sum_{i=1}^Nx_i \simeq\frac{1}{N}\sum_{i=1}^N\langle
x_i\rangle$; see Fig.~\ref{f1}. }

Naturally, (\ref{glo}, \ref{lo}) are to be compared with
the random supervising, where $N\rho$ supervised spins $\eta_i$ are
chosen completely randomly, i.e. without any dependence on $\y$.  Note
that besides (\ref{glo}, \ref{lo}) we also studied several other
supervising strategies, e.g. when $-1$ in the right-hand-side of
(\ref{glo}) or (\ref{lo}) is changed to $+1$. We shall not discuss such
strategies, since they proved to be sub-optimal; frequently they are
worse than the random supervising, and sometimes they are worse than
having no supervising at all. 

\subsection{Results}
\label{resu}

\subsubsection{Generation of images}

With a given coupling $J'$, we generated images $\x$ by Monte
Carlo simulation (Metropolis algorithm) from the two-dimensional Ising
model (\ref{eq:1}) on a square lattice with periodic boundary
conditions. The overall number of spins is $N=6400$ spins. Given the
noise probability $\epsilon$ (and hence $h$) from (\ref{eq:2},
\ref{calvin}), we flip each spin of the original image $\x$ generating
the noisy image $\y$.  The conditional averages in (\ref{eq:15},
\ref{eq:a50}, \ref{eq:a150}) are calculated by averaging over $2\times
10^3$ samples (we did check that this number suffices and the averages
saturate). 

\subsubsection{Three regimes of the square Ising ferromagnet}

Recall that the square (two-dimensional) Ising model (\ref{eq:1}) has
three regimes depending on the ferromagnetic coupling constant $J'$
\cite{landau}; see Fig.~\ref{f1}. For a small values of $J'$ the system
is in the paramagnetic state, where each spin $x_i$ is equally likely to
assume values $+1$ or $-1$. In the vicinity of a certain critical value
$J'_c$ we enter into the critical regime, where the paramagnetic state
gets unstable and there are long-range fluctuations (hence correlations)
\cite{landau}.  It is well-known that the infinite square lattice has a
second-order phase-transition point at $J'_c=\frac{1}{2}\ln
(1+\sqrt{2})\approx 0.440687$ \cite{landau}. This is close to the value
$J'_c=0.45$ observed in our numerics done on a finite lattice; see
Fig.~\ref{f1}. Another indication of the critical regime is that the
averaged magnetization $\frac{1}{N}\sum_{i=1}^N\overline{x}_i$ does not
coincide with the single-sample magnetization
$\frac{1}{N}\sum_{i=1}^N{x}_i$, as shown by Fig.~\ref{f1}. This is a
sign of strong-fluctuations. 

The third regime is set for a sufficiently large $J'$. Here fluctuations
are relatively small.  Indeed, Fig.~\ref{f1} shows that for $J'>0.46$ we
get that a single-sample behavior coincides with the averaged one. The
symmetry $x_i\to -x_i$ of the Hamiltonian (\ref{eq:1}) is spontaneously
broken (this process starts from the critical regime). Hence within each
given sample $\frac{1}{N}\sum_{i=1}^N{x}_i$ has a definite sign
\cite{landau}. 

Images generated in the third regime show pixels of one color on the
background of another color; see Fig.~\ref{f2}. Such figures are
interesting also because they are very susceptible to noise, and cannot
be recovered by standard (i.e. not active) methods; see Fig.~\ref{f2}.
In contrast, images generated in the critical regime do show an
interesting fine-grained structure. To some extent they can be recovered
via standard methods, though the active supervising still leads to a
serious improvement; see Figs.~\ref{f3} and \ref{f4}. 

\subsubsection{Images and overlaps}

The performance measure of strategies (\ref{glo}, \ref{lo}) is checked
via overlaps (\ref{eq:15}, \ref{eq:a150}). Overlap close to one is a
necessary condition for a good restoration. For definiteness, we
compared the performance (i.e. the overlap) of the present
non-supervised method with those of several standard filters|e.g. the
median filter or the Gaussian filter|that are employed in the
image-recognition; see e.g.  \cite{tanaka}. Numerical packages for
implementing these filters were taken from \cite{numeric}.  In agreement
with the fact that the considered method optimizes the overlap
(\ref{eq:15}), we found that these filtering methods produce a smaller
overlap than (\ref{eq:15}). 

We note that in the present situation we have a
possibility to look at additional performance measures: since we
generate the data ourselves|i.e. we have the original image $\x$|we can
employ directly an analogue of (\ref{pseudo}) that checks the restored
image (after supervising) with the original image $\x$: 
\BEA
\frac{1}{N(1-\rho)}\sum_{i=1}^N(1-n_i)\,\hat\xi_i' (\eee,\y)\,x_i,
\label{krolik}
\EEA
where $\xi_i' (\eee,\y)$ is defined in (\ref{eq:a50}). Note that since
we do not average over $\eee$ and $\y$, the overlaps (\ref{eq:a150}) and
(\ref{krolik}) are generally not equal for a considered finite value of
$N=6400$. We confirmed that (\ref{eq:a150}) and (\ref{krolik}) are not precisely
equal to each other, though they are normally quite close, since $N=6400$ is
still sufficiently large, and the self-averaging applies approximately.
In all relevant situations, if a strategy has a superior performance
according to (\ref{eq:a150}), then it is also superior according to
(\ref{krolik}). We also stress that all the results on differences
between the strategies were checked against varying the initial image
$\x$.

\begin{table*}
\begin{center}
\caption{ 
$\hat {\rm O}$ is the optimal overlap calculated via (\ref{eq:15})
without supervising for a square lattice with $N=6400$ spins and 
$N\rho$ supervised spins.  $\hat {\rm O}_{\rm R}$ refers to the random
supervising.  $\hat {\rm O}_{\rm G}$ and $\hat {\rm O}_{\rm L}$ refer
respectively to global and local strategies according to (\ref{glo}) and
(\ref{lo}); cf.~(\ref{eq:a14}). $\hat {\rm O}_{\rm MF}$ and $\hat {\rm O}_{\rm GF}$ 
are the overlaps obtained for (respectively) median filter and Gaussian filter.
abBoth filters were applied with the (minimal) radius $1$, i.e. via embedding each pixel into
a $3\times 3$ box \cite{book}.
For each set of parameters we underline the maximal overlap. 
\\
}
\begin{tabular*}{\textwidth}{@{}l*{15}{@{\extracolsep{0pt plus12pt}}l}}
\hline\hline
Parameters     &  $\hat {\rm O}$ &  $\hat {\rm O}_{\rm R}$ & $\hat {\rm O}_{\rm G}$ & $\hat {\rm O}_{\rm L}$ & $\hat {\rm O}_{\rm MF}$ & $\hat {\rm O}_{\rm GF}$ \\
\hline
$J'=0.500$, $\epsilon=0.3$, 
$\rho=0.1$~~~~~     &  0.90679 &  0.90727  & \u{0.92012} & 0.91850 & 0.90984 & 0.91563 \\
\hline
$J'=0.555$, 
$\epsilon=0.1$, 
$\rho=0.1$~~~~~     &  0.96341 &  0.96393  & \u{0.99213} & 0.99166 & 0.96313 & 0.96313 \\
\hline\hline
\label{tab1}
\end{tabular*}
\end{center}
\end{table*} 

\begin{table*}
\begin{center}
\caption{The same quantities as in Table~\ref{tab1}, but in the regime, where
$\hat {\rm O}_{\rm L}>\hat {\rm O}_{\rm G}$ and $\hat {\rm O}_{\rm L}>\hat {\rm O}_{\rm R}$. 
Note as well that $\hat {\rm O}_{\rm MF}<\hat {\rm O}_{\rm GF}$. \\ }
\begin{tabular*}{\textwidth}{@{}l*{15}{@{\extracolsep{0pt plus12pt}}l}}
\hline\hline
Parameters     &  $\hat {\rm O}$ &  $\hat {\rm O}_{\rm R}$ & $\hat {\rm O}_{\rm G}$ & $\hat {\rm O}_{\rm L}$ & $\hat {\rm O}_{\rm MF}$ & $\hat {\rm O}_{\rm GF}$  \\
\hline
$J'=0.455$, $\epsilon=0.1$, $\rho=0.09$~~~~~     &  0.88363 &  0.88616  & 0.92278 & \u{0.93314} & 0.83484 & 0.87094 \\
\hline
$J'=0.417$, $\epsilon=0.3$, $\rho=0.1$~~~~~     &  0.61359 &  0.64128  & {0.64830} & \u{0.66058} &  0.51936 & 0.53594 \\
\hline
$J'=0.417$, $\epsilon=0.3$, 
$\rho=0.2$~~~~~     & 0.61265 &  0.687619  & 0.682026 & \u{0.72048} & 0.51938 & 0.53594  \\
\hline\hline
\label{tab2}
\end{tabular*}
\end{center}
\end{table*} 

We identified two regimes in the behaviour of $\hat {\rm O}_{\rm L}$ and
$\hat {\rm O}_{\rm G}$, i.e. the overlaps (\ref{eq:a150}) within the local and global
strategies, respectively. In both regimes the maximal (better) of them is
larger than the overlap $\hat {\rm O}_{\rm R}$ obtained via the random
supervising at the same number $N\rho$ of supervised spins [see Tables 
\ref{tab1} and \ref{tab2}]:
\BEA
{\rm max}\left[\hat {\rm O}_{\rm L},\,\hat {\rm O}_{\rm G}\right]> \hat {\rm O}_{\rm R}.
\EEA

\begin{figure*}[htbp]
\centering
    \includegraphics[width=0.85\columnwidth]{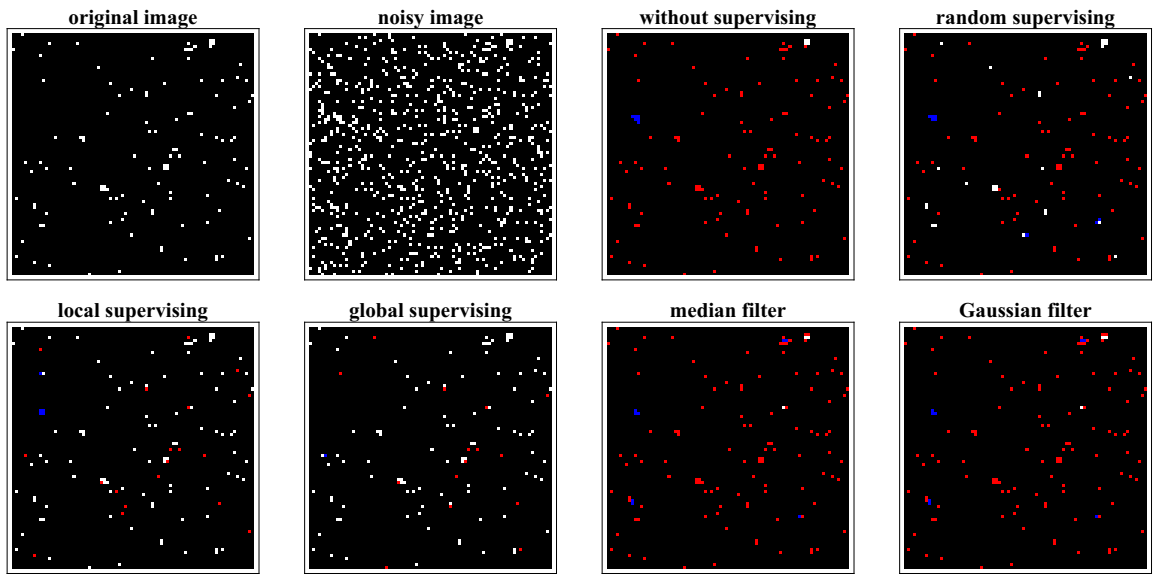}
\caption{ An example of active supervising for two scenarios (global and
local, as given by (\ref{glo}) and (\ref{lo}), respectively) under
$J'=0.555$, $\epsilon=0.1$ and $\rho=0.1$; see Table~\ref{tab1}. We
display the original image, its noise-corrupted version, the
non-supervised restoration and the active supervision via completely
randomly selected spins (pixels). On decoded images, we denote by red
(white) those white pixels of the original image that were decoded
wrongly (correctly). Likewise, blue (black) shows black pixels of the
original image that were decoded wrongly (correctly).\\ We also compare
with the (non-supervised) results obtained via two standard filters:
median and Gaussian. These filters were applied with the (minimal)
radius $1$, i.e. via embedding each pixel into a $3\times 3$ box
\cite{book}. Both active supervising scenarios recover the image
approximately, while other restoration methods are useless for this
example.  In this context, we emphasize again that a large overlap is
necessary but not sufficient for image recovery. } 
\label{f2}
\end{figure*}

Table~\ref{tab1} shows the first regime, where $J'$ is sufficiently
large, so that the system is away of the critical regime, which realized
for $J'\approx 0.45$; see Fig.~\ref{f1}. In these regime the strategy
(\ref{glo}) is dominant over (\ref{lo}), though their predictions are
close to each other:
\BEA
\hat {\rm O}_{\rm G}> \hat {\rm O}_{\rm L}, \qquad
\hat {\rm O}_{\rm G}\approx \hat {\rm O}_{\rm L}.
\label{shtap}
\EEA
Altogether, for $J'\geq 0.45$ the situation is close to the mean-field
regime.  This is additionally confirmed by the fact that the random
supervising does not provide any substantial improvement: $\hat {\rm
O}_{\rm R}\approx \hat {\rm O}$; see Table~\ref{tab1} and Fig.~\ref{f2},
and compare these with the discussion around (\ref{eq:35}).
Fig.~\ref{f2} describes regime (\ref{shtap}) and shows that without
supervising the real image restoration is absent, despite of the fact
that the overlap for this restoration regime is sizable; see
Table~\ref{tab1}. The reason for this is that the non-supervised
restoration is over-dominated by the prior information; cf. the discussion 
after (\ref{kio}). Fig.~\ref{f2} also shows that the image
restoration greatly improves after applying active methods. 
Both local and global methods lead to similar results here. 

\begin{figure*}[htbp]
\centering
\includegraphics[width=0.85\columnwidth]{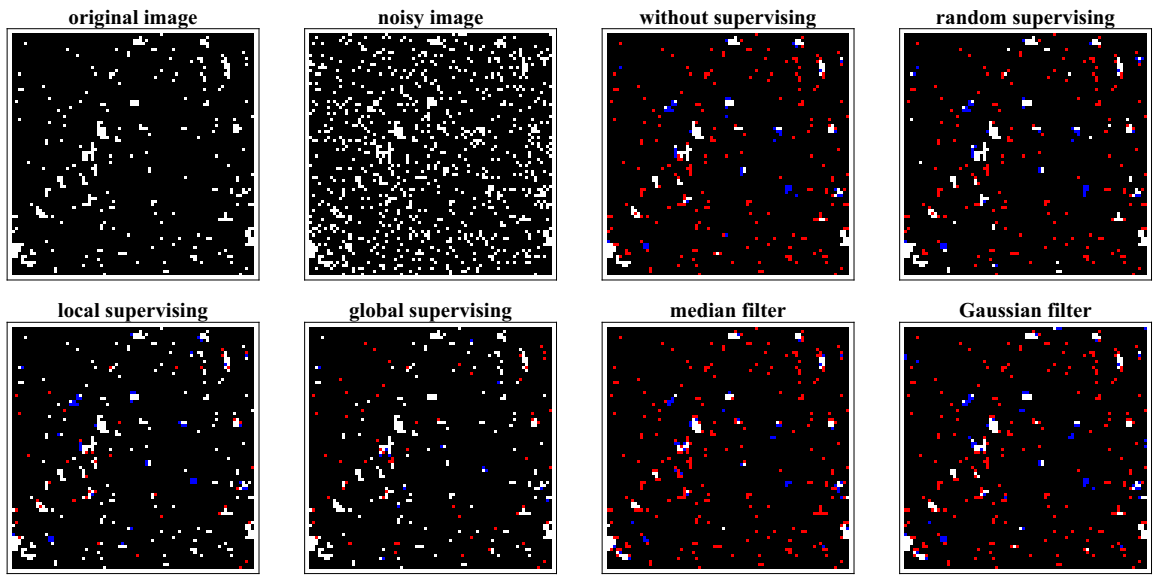}
\caption{ Various scenarios including active supervising (global and 
local, as given by (\ref{glo}) and (\ref{lo}), respectively) under
$J'=0.5$, $\epsilon=0.1$ and $\rho=0.1$; cf. Fig.~\ref{f2}. 
Both active supervising scenarios recover the
image, while the no-supervising and random supervising cases 
are useless. Notations coincide with those in Fig.~\ref{f2}.}
\label{f3}
\end{figure*}

\begin{figure*}[htbp]
\centering
\includegraphics[width=0.85\columnwidth]{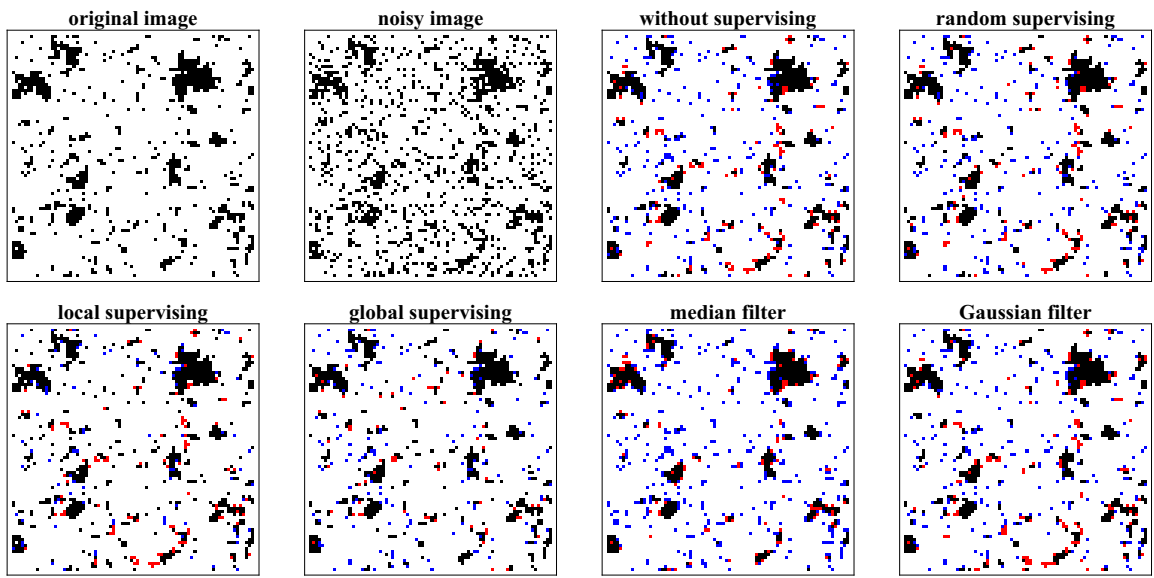}
\caption{Various scenarios including active supervising (global and 
local, as given by (\ref{glo}) and (\ref{lo}), respectively) under
$J'=0.455$, $\epsilon=0.1$ and $\rho=0.09$; cf. Figs.~\ref{f2} and \ref{f3}. 
Now no-supervising and random supervising cases are not useless, but 
fine details of the original image are recovered only after the local scenario. }
\label{f4}
\end{figure*}

For smaller values of $J'$, the local strategy is better than the global
one: $\hat {\rm O}_{\rm G}< \hat {\rm O}_{\rm L}$; see Table~\ref{tab2}.
Now the random supervising can improve over no supervising, and there
are cases (for a small $J'$), where the global strategy is worse than
the random: $\hat {\rm O}_{\rm G}< \hat {\rm O}_{\rm R}$. The reason why
the global strategy does not apply is clear, since for $J'<0.45$ the
mean-field method does not apply, in particular because the model is
within the crtical regime; see Fig.~\ref{f1}, where the critical regime
can be identified by the region, where $\sum_{i=1}^N\overline{x}_i$ and
$\sum_{i=1}^N\overline{y}_i$ sharply increase from zero to values larger
than $0.5$. It is encouraging that even for $J'\leq 0.45$ the local
strategy performs well, e.g. it is visibly better than the random
supervising; see Table~\ref{tab2}. Fig.~\ref{f3} illustrates this
situation for $J'=0.5$. It is seen that the performance of the
non-supervised restoration is still poor, although better than what was
seen for Fig.~\ref{f3}. Now the local active scenario is clearly better
than the global one. Finally, Fig.~\ref{f4} presents a representative
example of $J'=0.455$. Here all methods perform more or less reasonably,
but it is clearly seen that fine details of the original image are
captured only by the local supervising method. 

Thus we suggest that the local strategy (\ref{lo}) is to be applied for
this and similar models, because even when it is sub-optimal it is close
to the optimal strategy. 

\section{Numerical results with external magnetic field}
\label{magno}

So far we worked with the Hamiltonian (\ref{eq:3}) on a square lattice. 
The corresponding prior density $P(\x)\propto e^{-H(\x)}$ is generated by
the Ising Hamiltonian $H(\x)= - {J'}\sum_{\{i,k\}}x_ix_k$, which  
is symmetric with respect to the inversion $x_i\to -x_i$. This symmetry can be
broken with an external field $h_f$ that introduces a bias in the distribution of ${\bf x}$.
Now instead of (\ref{eq:3}) we look at
\BEA
\widetilde{H}(\x,\y)= - {J'}\sum_{\{i,k\}}x_ix_k -h_f\sum_i x_i -h\sum_i x_iy_i ,
\EEA
which leads to the joint probability $\widetilde P(\x,\y)\propto
e^{-\widetilde H(\x,\y)}$, and to the prior probability $\widetilde
P(\x)\propto e^{{J'}\sum_{\{i,k\}}x_ix_k+h_f\sum_i x_i }$. 

If $h_f$ assumes a small but generic value, the magnetization
$\frac{1}{N}\sum_{i=1}^N{x}_i$ and its observed value
$\frac{1}{N}\sum_{i=1}^N{y}_i$ become smoother functions of $J'$ that
are closer to their average values; see Fig.~\ref{fh1} and compare it
with Fig.~\ref{f1}. This is expected, because the second-order
phase-transition regime is now replaced by a smooth crossover from lower
to higher values of $\frac{1}{N}\sum_{i=1}^N{x}_i$ and
$\frac{1}{N}\sum_{i=1}^N{y}_i$; see Fig.~\ref{fh1}. Since the situation
is more stable (than for $h_f=0$), we can apply a smaller amount of
supervising (i.e. smaller values of $\rho$). 

\begin{figure*}[htbp]
\centering
    \includegraphics[width=0.85\columnwidth]{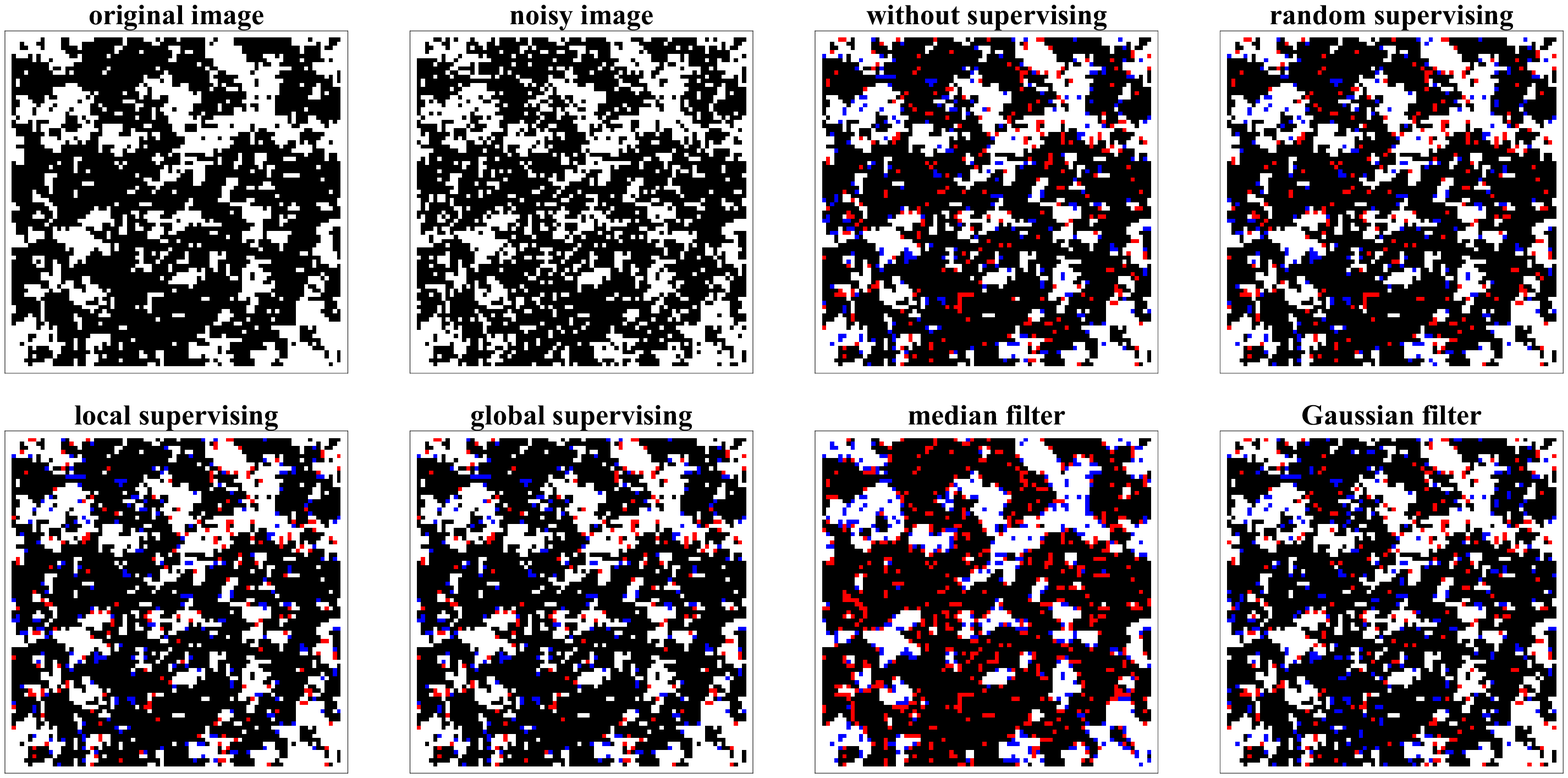}
\caption{ An example of active supervising for two scenarios (global and
local, as given by (\ref{glo}) and (\ref{lo}), respectively) under
$J'=0.39$, $h_f=0.01$ (non-zero magnetic field), $\epsilon=0.1$ and
$\rho=0.05$, i.e. everywhere we supervise $5\%$ of spins. Notations and
other parameters coincide with those in Fig.~\ref{f2}. } \label{fh2}
\end{figure*}

Fig.~\ref{fh2} shows the original and recovered images for a small but
generic value of the external field $h_f=0.01$ and for $J'=0.39$. It is
seen that the original image has a fine-grained structure. At $h_f=0.01$
such a structure is seen for $J'$ being roughly between $0.37$ and
$0.47$. As compared to other presented examples, here we applied a
smaller amount of supervising $\rho=0.05$ (i.e. the $5\%$ of spins is
supervised). Still this supervising is clearly useful, especially in its
active scenario. In Fig.~\ref{fh2} we are still far from the mean-field,
since $\hat{\rm O}_{\rm L}>\hat{\rm O}_{\rm G}$; cf. Table \ref{tab5}
for further data.

\begin{figure}[!htb]
\centering
  \subfigure[]{
    \includegraphics[width=0.45\columnwidth]{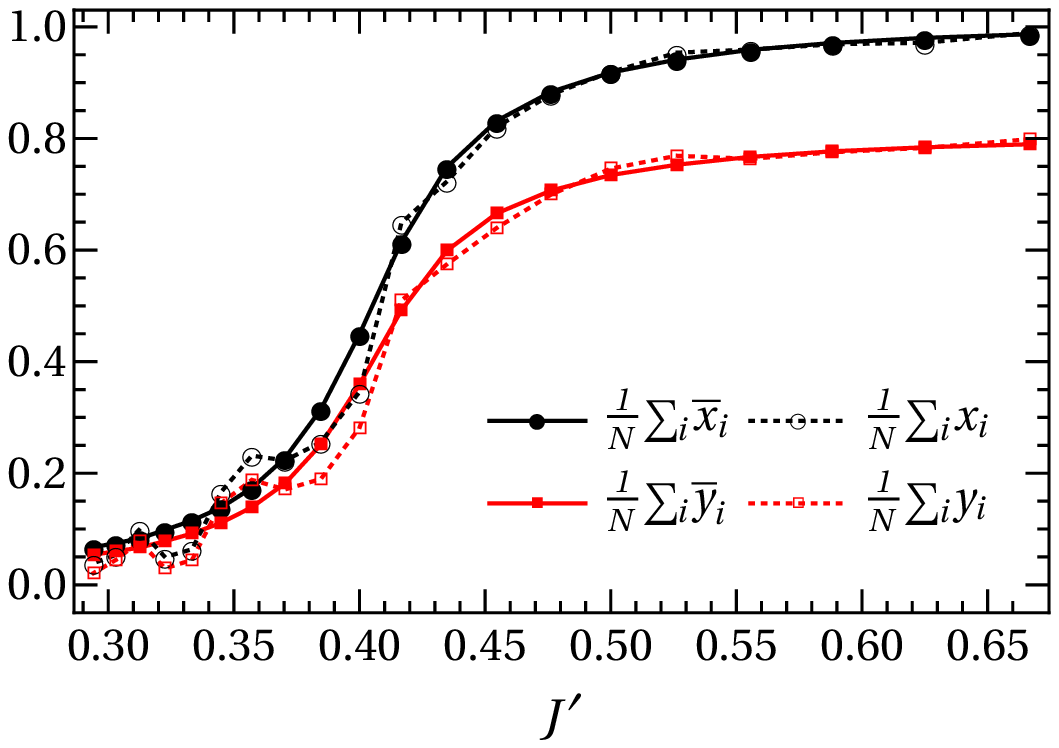}
\label{fig1a}
    } 
    \subfigure[]{
    \includegraphics[width=0.45\columnwidth]{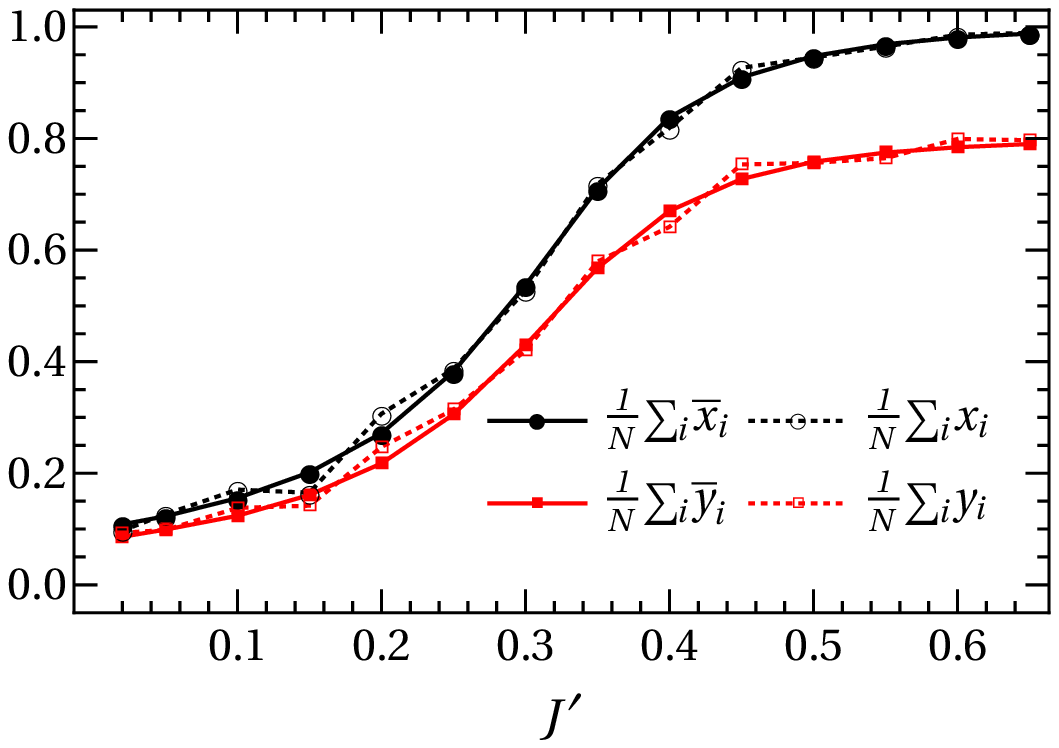}
    }
    \caption{ (a)  The same as in Fig.~\ref{f1}, but with an external magnetic field. 
Magnetization $\frac{1}{N}\sum_{i=1}^N{x}_i$ and its
observed value $\frac{1}{N}\sum_{i=1}^N{y}_i$ versus $J'$ under a fixed
$\epsilon=0.1$ in (\ref{eq:2}) (hence $h=1.09861$), and an external magnetic field $h_f=0.01$. 
Single realizations of $\x=\{x_i\}_{i=1}^N$ and $\y=\{y_i\}_{i=1}^N$ 
(denotted by dotted lines) are closer to the average behavior than in Fig.~\ref{f1}. 
\\ (b) The same as in (a), but for $h_f=0.1$. It is seen that single realizations of 
$\x=\{x_i\}_{i=1}^N$ and $\y=\{y_i\}_{i=1}^N$ converge to the their averages upon increasing $h_f$.   }
\label{fh1}
\end{figure}

\comment{
\begin{figure*}[htbp] \centering

\caption{ The same as in Fig.~\ref{f1}, but with an external magnetic field. 
Magnetization $\frac{1}{N}\sum_{i=1}^N{x}_i$ and its
observed value $\frac{1}{N}\sum_{i=1}^N{y}_i$ versus $J'$ under a fixed
$\epsilon=0.1$ in (\ref{eq:2}) (hence $h=1.09861$), and an external magnetic field $h_f=0.01$. 
Now single realizations of $\x=\{x_i\}_{i=1}^N$ and $\y=\{y_i\}_{i=1}^N$ 
(denotted by dotted lines) always coincide with the average behavior. }
\label{fh1}
\end{figure*}
}

\begin{table*}
\begin{center}
\caption{We present values for overlaps for external magnetic field
$h_f=0.01$ a square lattice with $N=6400$ spins and $N\rho$ supervised
spins: $\hat {\rm O}$ (the optimal overlap calculated via (\ref{eq:15})
without supervising), $\hat {\rm O}_{\rm R}$ (random supervising), $\hat
{\rm O}_{\rm G}$ and $\hat {\rm O}_{\rm L}$ refer respectively to global
and local strategies according to (\ref{glo}) and (\ref{lo});
cf.~(\ref{eq:a14}). $\hat {\rm O}_{\rm MF}$ and $\hat {\rm O}_{\rm GF}$
are the overlaps obtained for (respectively) median filter and Gaussian
filter.\\ For the global strategy the fraction of supervised spins is
$\rho_{\rm G}=0.1$. For the local strategy this value $\rho_\L$ is
generally lower, as shown below. Despite of this fact the local strategy
produced better overlaps: $\hat {\rm O}_{\rm L}>\hat {\rm O}_{\rm G}$. 
}
\begin{tabular}{ccccccc}
\hline\hline
Parameters     &  $\hat {\rm O}$ &  $\hat {\rm O}_{\rm R}$ & $\hat {\rm O}_{\rm G}$ & $\hat {\rm O}_{\rm L}$ & $\hat {\rm O}_{\rm MF}$ & $\hat {\rm O}_{\rm GF}$ \\
\hline
$J'=0.38$, $\epsilon=0.1$, 
$\rho_\L=0.05$~~~~~     &  0.82766 &  0.83174 & 0.82865 & \u{0.85987} &
0.67563 & 0.82250 \\
\hline
$J'=0.39$, 
$\epsilon=0.1$, 
$\rho_\L=0.05$~~~~~     &  0.82578 &  0.830263 & 0.82622 & \u{0.85691} &
0.69625 & 0.82375 \\
\hline
$J'=0.40$, $\epsilon=0.1$, 
$\rho_\L=0.07$~~~~~     &  0.84375 &  0.84543 & 0.84688 & \u{0.88542} &
0.75516 & 0.83188 \\
\hline
$J'=0.41$, $\epsilon=0.1$, 
$\rho_\L=0.07$~~~~~     &  0.84891 &  0.85719 & 0.85208 & \u{0.89147} &
0.76359 & 0.84375 \\
\hline
$J'=0.42$, $\epsilon=0.1$, 
$\rho_\L=0.08$~~~~~     &  0.85406 &  0.85564 & 0.86701 & \u{0.90353} &
0.78313 & 0.84094 \\
\hline
$J'=0.43$, $\epsilon=0.1$, 
$\rho_\L=0.09$~~~~~     &  0.86688 &  0.86058 & 0.86597 & \u{0.91896} &
0.80328 & 0.85593 \\
\hline
$J'=0.44$, $\epsilon=0.1$, 
$\rho_\L=0.1$~~~~~     &  0.87813 &  0.89063 & 0.87951 & \u{0.93611} &
0.83375 & 0.87344\\
\hline
$J'=0.45$, $\epsilon=0.1$, 
$\rho_\L=0.1$~~~~~     &  0.90047 &  0.90104 & 0.90381 & \u{0.95243} &
0.86953 & 0.89188 \\
\hline\hline
\label{tab5}
\end{tabular}
\end{center}
\end{table*}

\section{Summary}
\label{sumo}

Restoration of noise-corrupted images is a fundamental problem of
statistics, and it is also of obvious practical importance
\cite{geman,besag,scot,tanaka,cohen,nishimori,nishi}. It joins together
mathematical and physical statistics, since the simplest non-trivial
approach to this problem is based on the two-dimensional, square-lattice
Ising model with a ferromagnetic interaction between neighbouring binary
pixels (spins) \cite{geman,scot}. The prior probability of images within
this model is determined via the Gibbs distribution, where the
ferromagnetic coupling constants account for natural smoothness
(correlation) between neighbouring spins. Hence the Ising model was
extensively studied in the context of image restoration
\cite{geman,besag,scot,tanaka,cohen,nishimori,nishi}. The optimal
approach to restoration within the Ising model is based on maximizing
the average overlap (the inverse Bayesian risk). This approach demands
calculating Gibbsian (or equilibrium) averages of all Ising spins. In
the sense of the average overlap, this approach outperforms those based
on various types of filtering (e.g. median filtering) \cite{tanaka}.
However, filtering approaches are easier to implement in practice. 

Here we studied a ferromagnetic Ising model for active restoration of
images, where prior information on certain (supervised) spins is
requested using the initial observation of the noise-corrupted image.
For a given number of supervised spins, the optimal supervising strategy
maximizes the average overlap of non-supervised spins with their true
(requested) values. Generally, this optimal strategy is not known. We
found the optimal supervising strategy within the mean-field
approximation, which is introduced via letting all spins couple with
each other. It applies to realistic (two-dimensional) images, whenever
the inter-spin coupling is large (the image is smooth) and hence
fluctuations are small. The optimal strategy in this situation amounts
to supervising those pixels (spins) that do not agree with the sign of
the overall magnetization.  The strategy shows that there is an upper
limit on the number of supervised spins, beyond of which it is
meaningless to supervise, since the performance will not change. 

The mean-field approximation allows to study the active supervision in
detail. It also leads to strategies that can apply more generally, i.e.
in the critical regime of the two-dimensional model, where correlations
are long-range and probability distributions are self-similar
\cite{bialek}. The critical regime applies to natural images
\cite{ruderman}. We explored one such strategy, where we supervise only
those spins (pixels) that disagree with their Bayesian estimates. The
performance of this strategy is comparable with the optimal one where
the latter is known, i.e. the mean-field applies.  Otherwise (e.g. in
the critical regime), it outperforms over all other methods we tried. 

In this study we assumed that hyperparamaters of the model|the
smoothness parameter $J$ of the prior probability (ferromagnetic
coupling), the bias $h_f$ (external magnetic field) and the noise
parameter $h$|are known. Generally, this is not the case:
hyperparameters are to be determined from data.  This is one instance of
the hyperparameter learning for the Ising model that recently attracted
much attention \cite{hyper_naive_mf,hyper}; see \cite{hyper_review,berg}
for reviews. In Appendix we studied the hyperparameter learning within
the mean-field approximation, and saw that this problem is non-trivial
due to non-identifiability: if (for $h_f=0$ taken to be zero for
simplicity) both hyperparameters $J$ and $h$ are unknown, then neither
of them can be found via the standard maximum likelihood approach. The
non-identifiability problem is crucial for image restoration, since
imprecisely known hyperparameters may lead to a larger value of the
average overlap (over-confidence), thereby creating a false impression
about the performance of the restoration.  Appendix also shows that
within the mean-field approximation the non-identifiability problem can
be resolved via the active supervising. A pertinent open problem is how
the active supervising alters hyperparameter learning for the
two-dimensional situation, especially in its critical regime.

\section*{Acknowledgement}

A.E.A.~acknowledges discussions with A.~Galstyan. A.E.A. was supported 
by SCS of Armenia (Grant No.~18RF-015). This work is supported 
by National Natural Science Foundation of China (Grant No.~11505071), the 
Programme of Introducing Talents of Discipline to Universities under Grant
No.~B08033 and the Fundamental Research Funds for the Central Universities.

\appendix

\section*{Appendix: Unknown hyperparameters}
\label{hypo}

\subsection{The maximum-likelihood method for estimation of hyperparameters}

The standard maximum-likelihood method of estimating unknown
hyper-parameters is to start with the observations $\y$ and
probabilities $P^\circ(\y)$ at trial values of the hyper-parameters
\cite{ephraim_review}. For the mean-field model studied in section \ref{mf} these
trial values are $J^\circ$ and $h^\circ$, while the correct values of
hyperparameters are still denoted by $J$ and $h$. 

Now if $N\gg 1$, then $\ln P^\circ(\y)$ self-averages:
\begin{eqnarray}
  \label{eq:17}
  \ln P^\circ(\y)\simeq {\sum}_{\y}P(\y) \ln
  P^\circ(\y),
\end{eqnarray}
with the true (i.e. containing the correct hyper-parameters $J$ and
$h$) probability $P(\y)$. The global maximum of (\ref{eq:17}) is
reached (among other possibilities) at $J^\circ=J$ and $h^\circ=h$
\cite{ephraim_review}. This fact follows from the positivity of the
relative entropy: $\sum_{\y}P(\y) \ln\frac{P(\y)}{ P^\circ(\y)}\geq
0$. However, the maximum is generally not unique, so the maximization
over $J^\circ$ and $h^\circ$ does not need to return true values $J$ and
$h$. In practice the global optimization is generally intractable;
hence it is implemented via an expectation--maximization (EM) method
(Baum-Welch algorithm) \cite{ephraim_review}.

Now we should use (\ref{rashid}, \ref{eq:10}) with $J$, $h$ and $m$
replaced by respectively $J^\circ$, $h^\circ$ and $m^\circ$, where 
instead of (\ref{eq:11}) we get
\begin{eqnarray}
 m^\circ =\frac{1}{N}\sum_i \tanh(J^\circ m^\circ+h^\circ y_i)
= \sum_{y=\pm 1}\pi(y)\tanh(J^\circ m^\circ+yh^\circ ).
  \label{eq:2121}
\end{eqnarray}
Note that $ m^\circ$ self-averages with probability $\pi(y)$ 
containing true values of $J$ and $m$. 

Due to factorization (\ref{eq:10}) the maximization of
(\ref{eq:17}) amounts to maximizing
\begin{eqnarray}
  \label{eq:19}
\sum_{y=\pm 1}\pi(y) \ln \pi^\circ(y),
\end{eqnarray}
where $\pi(y)$ is given by (\ref{eq:10}), and where [cf.~(\ref{eq:9.1})]
\begin{eqnarray}
  \label{eq:21}
\pi^\circ(y)=  \frac{1}{2}[1+y\tanh(J^\circ m^\circ)\tanh(h^\circ)],
\end{eqnarray}

Now it is clear from (\ref{eq:10}, \ref{eq:21}) that for $J>1$
\footnote{For $J<1$ we get $\pi(1)=\pi(-1)=1/2$. Then (\ref{eq:2121})
  leads to $m^\circ=0$. This satisfies (\ref{eq:20}), but there is no
  constraint on $J^\circ$ and on $h^\circ$, i.e. no determination of
  hyperparameters is possible whatsoever. This is expected, because
  for $J<1$ the initial distribution of the spins is completely
  unbiased. } the maximization of
(\ref{eq:19}) will lead to $\pi(1)=\pi^\circ(1)$, and this global
maximum is achieved for
\begin{eqnarray}
  \label{eq:20}
\tanh(Jm)\tanh(h)=
\tanh(J^\circ m^\circ)\tanh(h^\circ).
\end{eqnarray}
Using in (\ref{eq:2121})
$\tanh[a+b]=\frac{\tanh[a]+\tanh[b]}{1+\tanh[a]\tanh[b]}$ and
(\ref{eq:20}) we simplify (\ref{eq:2121}) as follows
\begin{eqnarray}
  \label{eq:22}
  m^\circ=\tanh[J^\circ  m^\circ].
\end{eqnarray}
Eq.~(\ref{eq:20}) cannot determine two unknowns $J^\circ$ and
$h^\circ$. If one of them is known precisely, e.g. $h=h^\circ$, the
other one will be found via (\ref{eq:20}). But if both $J^\circ$ and
$h^\circ$ are unknown, neither of them will be found. This is an
example of the non-identifiability problem in inference
\cite{1957,ito}.

\subsection{Overlap and over-confidence}

The overlap and magnetization still self-average under the correct
values $J$ and $h$ [cf.~(\ref{eq:13})]:
\begin{eqnarray}
  \label{eq:18}
  \O^\circ = \sum_{y=\pm 1}\pi(y)\tanh[\,|\, J^\circ m^\circ+yh^\circ |\,],
\end{eqnarray}
where $\pi(y)$ is given by (\ref{eq:10}), and we employed $\hat
\xi_i^\circ (y)={\rm sign}[J^\circ m^\circ+h^\circ y_i]$. With 
(\ref{eq:20}) and (\ref{eq:22}) one deduces from (\ref{eq:18}):
\begin{eqnarray}
  \label{eq:1818}
  \O^\circ = {\rm max}\left(\,\tanh[\, J^\circ |m^\circ|\,], \,
\tanh[\, h^\circ \,]\,\right).
\end{eqnarray}
Recall that $J^\circ$ and $h^\circ$ hold (\ref{eq:20}). Under this constraint $
\O^\circ$ in (\ref{eq:1818}) can be either larger or smaller than the
optimal overlap $\hat\O$ given by (\ref{eq:13}); e.g. one can take
$J^\circ$ sufficiently large and $h^\circ$ small so as to hold
(\ref{eq:20}) and get $\O^\circ>\hat\O$ from (\ref{eq:1818}). Hence
the overlap is a good criterion for distinguishing the optimal
solution only if the hyperparameters are known precisely.

Once hyperparameters cannot be found precisely, there can be two
possibilities for the overlap under trial values $\O^\circ$ [see
(\ref{eq:20}, \ref{eq:1818})] and the optimal overlap $\hat \O$ given
by (\ref{eq:13}): $\O^\circ<\hat\O$ and $\O^\circ>\hat\O$. The former
is a fair situation, where imprecise knowledge of hyperparameters
($J^\circ\not =J$ and $h^\circ\not =h$) brings in a reduction in the
overlap. In contrast, the latter is a dangerous situation, where
imprecise knowledge leads to over-confidence (over-fitting). We feel
that so far not enough attention was devoted to the over-confidence
phenomenon both computationally and theoretically. One of the
advantages of the present model is that it makes the phenomenon
obvious.

To avoid over-confidence, one can minimize $\O^\circ$ in
(\ref{eq:1818}) over $J^\circ$ and over $h^\circ$ under constraint
(\ref{eq:20}). Clearly, this procedure will lead to $\O^\circ<\hat\O$.
For our situation this produces:
\begin{eqnarray}
  \label{eq:4}
  \O^\circ=\sqrt{\tanh(Jm) \tanh(h)}\leq \hat\O, \\
\tanh(J^\circ m^\circ)= \tanh(h^\circ)=\sqrt{\tanh(Jm) \tanh(h)}.
  \label{beli}
\end{eqnarray}
Eq.~(\ref{beli}) is a reasonable setting for unknown
hyperparameters. This method works theoretically, but its algorithmic
implementation in hyperparameter learning algorithms is not yet clear.

\subsection{Active supervising and hyper-parameter learning}

We shall now study how the active supervising|which was designed so
as to increase the overlap|influences on the determination of the
hyper-parameters. Instead of (\ref{eq:17}) and (\ref{eq:19}) we should
now maximize the following (self-averaged) expression over the trial
hyper-parameters $h^\circ$ and $m^\circ$:
\begin{eqnarray}
  \label{eq:180}
\sum_{\eee,\y,\n}P(\eee,\y)P(\n|\y)\ln
[P^\circ(\eee,\y)P(\n|\y)], 
\end{eqnarray}
where we note that $P(\n|\y)$ does not depend on trial
hyperparameters, since it is designed on the ground of observations
$\y$. When the supervising is absent, i.e. $P(\n|\y)=1$ for
$\n=(0,...,0)$, (\ref{eq:180}) reverts to 
(\ref{eq:17}).

We now additionally assume that 
\begin{eqnarray}
  \label{cora}
P(\n|\y)=\prod_{k=1}^N p(n_k|y_k).
\end{eqnarray}
Using (\ref{rashid}, \ref{eq:10}, \ref{cora}) we write:
\begin{eqnarray}
  \label{eq:6}
&&  P^\circ(\eee,\y)P(\n|\y)=\prod_{k=1}^N \chi(\eta_k,y_k,n_k), \\
&&  \chi(\eta,y,n) =\pi^\circ(\eta,y)p(n|y)\delta_{n1}+
  \pi^\circ(y)p(n|y)\delta_{n0}, 
\end{eqnarray}
where $\delta_{n0}$ is the Kronecker's delta.  Analogous formula can
be written for $P(\eee,\y)P(\n|\y)$. Thus maximizing (\ref{eq:180})
over trial hyperparameters $J^\circ$ and $h^\circ$ amounts to
maximizing
\begin{eqnarray}
\label{shish}
&&\sum_{\et, y}p(1|y)\,\pi(\et,y) \ln[\,p(1|y)\pi^\circ(\et,y)\,] +  \sum_{y}p(0|y)\,\pi(y) \ln[\,p(0|y)\pi^\circ(y)\,] \\
&=&\sum_{\et, y}p(1|y)\,\pi(\et,y) \ln[\,p(1|y)\,] +
\sum_{\et, y}p(1|y)\,\pi(\et,y) \ln[\,\pi^\circ(\et,y)/\, \pi^\circ(y)\,]  \nonumber\\ 
& & +\sum_{\et, y}p(1|y)\,\pi(\et,y) \ln[\,\pi^\circ(y)\,] 
+\sum_{y}p(0|y)\,\pi(y) \ln[\,\pi^\circ(y)\,] + 
\sum_{y}p(0|y)\,\pi(y) \ln[\,p(0|y)\,] \nonumber\\   \label{eq:25}
&=&\sum_{\et, y}p(1|y)\,\pi(\et,y) \ln[\,\pi^\circ(\et,y)/ \pi^\circ(y)\,]+
\sum_{y}\,\pi(y) \ln[\,\pi^\circ(y)\,]\\
& & +\sum_{ y}p(1|y)\,\pi(y) \ln[\,p(1|y)\,] + \sum_{y}p(0|y)\,\pi(y) \ln[\,p(0|y)\,].
\label{nav}
\end{eqnarray}
Thus we can maximize only (\ref{eq:25}), since (\ref{nav}) does not depend on trial hyperparameters.
Note that the last term in (\ref{eq:25}) is the expression
(\ref{eq:19}) without supervising; it is recovered from (\ref{eq:25})
for $p(1|y)=0$. The first term in (\ref{eq:25}) is the self-averaged
expression for $\ln[\,\pi^\circ(\et|y)]$.

Inspecting directly (\ref{shish}) or (\ref{eq:25}) we conclude that now
the maximization over $J^\circ$ and $h^\circ$ will recover the true
values of the hyper-parameters: $J^\circ=J$ and $h^\circ=h$, even in the
regimes (\ref{yu}, \ref{la}), where the overlap is optimized.  Note that
this conclusion holds even for the random supervising (\ref{eq:35}),
where its origin is especially clear: together with $\sum_{y}\pi(y)
\ln\pi^\circ(y)$ we maximize $\sum_{\et, y} \pi(\et,y)
\ln[\pi^\circ(\et,y)]$, and the later maximization leads to $J^\circ=J$
and $h^\circ=h$, which also maximizes the former; cf.~(\ref{eq:20}).

\end{document}